\documentclass[pra,twocolumn,showpacs]{revtex4}
\usepackage{graphicx}
\begin{document}

\bibliographystyle{unsrt}

\title{Violation of Bell's Inequality with Photons from Independent Sources}
\author{T.B. Pittman and J.D. Franson}
\affiliation{Johns Hopkins University,
Applied Physics Laboratory, Laurel, MD 20723}

\date{\today}

\begin{abstract}
We report a violation of Bell's inequality using one photon from a parametric down-conversion source and a second photon from an attenuated laser beam.  The two photons were entangled at a beam splitter using the post-selection technique of Shih and Alley [Phys. Rev. Lett. {\bf 61}, 2921 (1988)].  A quantum interference pattern with a visibility of 91\% was obtained using the photons from these independent sources, as compared with a visibility of 99.4\% using two photons from a central parametric down-conversion source. 
\end{abstract}


\maketitle

Nearly all experimental tests of nonlocality based on Bell's inequality \cite{bellbook} have used pairs of photons emitted in an entangled state by a common source \cite{aspect81,kwiat95}.  Nonetheless, Yurke and Stoler have shown that Bell's inequalities can be violated even if the two particles do not originate from a common source \cite{yurke92}, and this has been demonstrated \cite{jennewein02,zhao03} using two pairs of entangled photons from parametric down-conversion combined with entanglement swapping \cite{zukowski93}.  Here we describe an experimental violation of Bell's inequality using one photon from parametric down-conversion and a second photon from an attenuated laser beam.  The two photons were entangled at a beam splitter using the post-selection technique of Shih and Alley \cite{shih88}.  The ability to obtain non-classical interference effects using photons from independent sources is an important requirement for an optical approach to quantum information processing \cite{knill01,franson02}.

One of the interesting features of this experiment is the fact that a coherent state produced by a laser is essentially a classical beam of light.  Nevertheless,  quantum interference patterns with visibilities as high as 91\% were obtained using the two photons from independent sources.  For comparison, a visibility of 99.4\% was obtained using two photons from a central parametric down-conversion source.  The lower visibility for photons from independent sources was due to imperfect mode matching and a decreased signal-to-noise ratio, as will be discussed in more detail below.

\begin{figure}[b]
\includegraphics[angle=-90,width=3in]{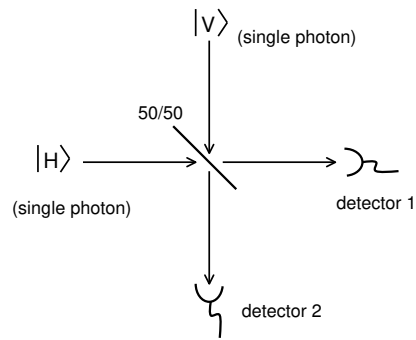}
\vspace*{-.35in}
\caption{An overview of the Shih-Alley technique \protect\cite{shih88} used to violate Bell's inequalities.  In our experiment, one of the input photons is heralded from a down-conversion pair, while the second input photon is post-selected from an auxiliary weak coherent state.}
\label{fig:overview}
\end{figure}

In the first Bell-inequality experiment to use photons from parametric down-conversion, Shih and Alley \cite{shih88} combined two photons of a down-conversion pair at a 50/50 beam splitter as shown in Figure \ref{fig:overview}. In this experiment,   
one of the photons is horizontally polarized (denoted $|H\rangle$), while the second photon is vertically polarized ($|V\rangle$).  Provided these photons are otherwise indistinguishable \cite{feynmanlectures}, the output state can be expressed as $\frac{1}{2}(|H_{1}V_{2}\rangle + i|H_{1}V_{1}\rangle + i|H_{2}V_{2}\rangle - |V_{1}H_{2}\rangle)$, where the subscripts denote the output port of the beam splitter. Coincidence measurements in which one photon is detected in each output port post-selects an entangled state of the form:

\begin{equation}
|\psi^{-}\rangle = \frac{1}{\sqrt{2}}(|H_{1}V_{2}\rangle - |V_{1}H_{2}\rangle)
\label{eq:sa}
\end{equation}

\noindent which is suitable for a test of Bell's inequalities \cite{reid86}.

In our experiment, one of the inputs to the beam splitter in Figure \ref{fig:overview} was a single-photon heralded from a down-conversion pair \cite{hong86}, while the second input photon was derived from a weak coherent state.  The quality of the resulting post-selected state (\ref{eq:sa}), and consequently the ability to violate Bell's inequality in this situation, relied on the indistinguishability of the photons from these two different sources \cite{ou97}.

There have been several experiments demonstrating various aspects of the potential indistinguishability of down-converted photons and photons from weak coherent states (see, for example, \cite{koashi96,kuzmitch00,resch02,lvovsky}).
In particular, Rarity and Tapster performed an experiment \cite{rarity97} in which the well known Hong-Ou-Mandel ``dip'' \cite{hong87} was observed when a weak coherent state was mixed with a single heralded down-conversion photon at a 50/50 beam splitter. The key to their experiment was the elimination of timing information that, in principle, could distinguish the detected photons. This was accomplished by using ultrashort laser pulses to pump the down-conversion source, followed by narrowband interference filters to increase the coherence length of the photons and produce an overlap of their wavepackets \cite{zukowski95,ou97}. This same technique has also been successfully used in several experiments demonstrating higher-order interference effects involving single photons emitted from multiple down-conversion events
\cite{jennewein02,zhao03,bouwmeester,pan98,ou99,linares01,pan01,deried02,yam03}.
 
\begin{figure}
\hspace*{-.25in}
\includegraphics[angle=-90,width=3.75in]{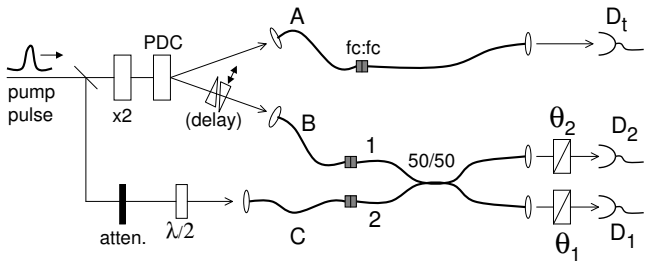}
\vspace*{-1.25in}
\caption{A simplified schematic of the experiment used to post-select two-photon entanglement from independent sources. Details and symbols are described in the text.}
\label{fig:experiment}
\end{figure}

In the first step of our experiment, the observation of a high visibility dip \cite{rarity97} signalled the experimental conditions necessary for the required indistinguishability of the independent photon sources.  The second step then involved rotating the polarizations to repeat the Shih-Alley experiment \cite{shih88} depicted in Figure \ref{fig:overview}.  A simplified schematic of our experimental apparatus is shown in Figure \ref{fig:experiment}.

Short laser pulses ($\approx$150 fs) at 780nm from a mode-locked Ti:Sapphire laser were frequency doubled in a BBO crystal (labelled x2), providing UV pulses (390nm) that were then used to pump a second BBO crystal (labelled PDC) optimized for degenerate type-I non-collinear parametric down-conversion. This down-conversion source emitted pairs of horizontally polarized photons at 780nm which were then coupled into single mode optical fibers labelled $A$ and $B$.  An optical delay unit formed by two translating glass wedges was inserted in one of the free-space down-conversion beams.

A small fraction of the original 780nm pumping beam was picked off and used as the weak coherent state.  A variable attenuator was used to reduce the magnitude of this coherent state to the single-photon level, and a half-wave plate ($\lambda/2$) was used to rotate the linear polarization state as needed.  Photons from the weak coherent state were coupled into the single-mode fiber labelled $C$.

Fibers $B$ and $C$ were connected to the input ports of a fused 3dB fiber coupler (labelled 50/50) that served as the 50/50 beam splitter for the Shih-Alley experiment of Figure \ref{fig:overview}.  The output fibers of the 3dB coupler were used to direct the output beams to two single-photon detectors $D_{1}$ and $D_{2}$.  These detectors, as well as the single-photon triggering detector $D_{t}$, were preceded by narrowband interference filters (not shown) centered at 780nm.  $\theta_{1}$ and $\theta_{2}$ were polarization analyzers.

\begin{figure}[b]
\vspace*{-.25in}
\hspace*{-.2in}
\includegraphics[angle=-90,width=3.5in]{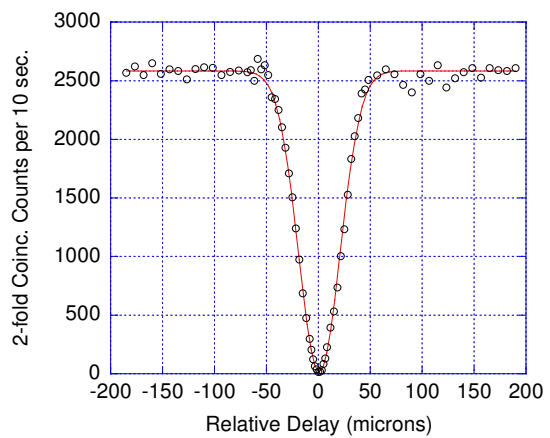}
\vspace*{-.25in}
\caption{Experimental results of a two-photon Hong-Ou-Mandel test \protect\cite{hong87} using the apparatus of Figure \protect\ref{fig:experiment}. The data shows coincidence counts between detectors $D_{1}$ and $D_{2}$ as a function of relative optical delay between the down-converted beams when fibers $A$ and $B$ were used as the inputs to the 50/50 beam splitter. The solid line is a least-squares fit to a simple Gaussian function with a visibility of ($99.4 \pm 0.1$)\%.}
\label{fig:homdip}
\end{figure}

As shown in Figure \ref{fig:experiment}, the various fibers were joined with standard fc:fc connectors.  
In order to test the quality of the 3dB fiber coupler, along with the alignment and correlations of the down-converted beams, we first repeated the Hong-Ou-Mandel experiment \cite{hong87} by temporarily connecting fibers $A$ and $B$ (rather than $B$ and $C$) to the two input ports ($1$ and $2$) of the coupler.  $\theta_{1}$ and $\theta_{2}$ were set to their horizontal values to match the horizontal polarizations of the down-converted photon pairs, and the effects of birefringence in the single-mode fibers were minimized using standard fiber polarization controllers.  Using interference filters with a 10nm FWHM bandpass, coincidence counts between $D_{1}$ and $D_{2}$ were recorded as a function of the relative optical delay imposed by the glass wedges. As shown in Figure \ref{fig:homdip}, a standard two-photon dip was observed with a visibility of ($99.4 \pm 0.1$)\%. This high visibility indicated a nearly perfect 50/50 beam splitter and a minimum of scattered photons from the original 780nm pumping pulses; both of these were critical requirements for the subsequent experiments of interest.

For the subsequent experiments, the fibers were reconnected as shown in Figure \ref{fig:experiment}. The detection of a down-converted triggering photon by $D_{t}$ heralded the presence of the horizontally polarized twin photon in fiber $B$ with some limited probability, and was also used to gate the coincidence counting between $D_{1}$ and $D_{2}$ \cite{rarity97}. Because the probability of two down-conversion events from a single pump pulse was negligibly small, a gated coincidence count between $D_{1}$ and $D_{2}$ therefore implied (with high probability) the joint detection of the heralded single-photon and a single photon from the vertically polarized weak coherent state coupled into fiber $C$.  In these instances the three-photon (ie. gated two-photon) coincidence measurements  post-selected an entangled state as in the Shih-Alley experiment \cite{shih88}, but with independent sources as depicted in Figure \ref{fig:overview}. 

The allowable magnitude of the weak coherent state was limited by the probability of successfully heralding a single photon from a down-conversion pair.  Roughly speaking, the probability of a three-photon event of interest $P$ was proportional to $\Gamma\alpha$, where $\Gamma$ corresponds the probability (per pump pulse) of a detectable down-conversion pair, and $\alpha$ is the probability per pulse of a detectable single photon from the weak coherent state. On the other hand, the largest background noise contribution $P'$ (ie. unwanted three-photon detection event) was proportional to $H\Gamma\alpha^{2}$, where $H\Gamma$ denotes the probability of detecting a down-converted triggering photon while the twin photon has been lost.  In order to observe the high visibility gated two-photon interference effects necessary for a violation of Bell's inequality, we require that $P \gg P'$ (eg. high signal-to-noise), which implies that $\alpha \ll \frac{1}{H}$.

\begin{figure}[b]
\vspace*{-.15in}
\hspace*{-.2in}
\includegraphics[angle=-90,width=3.5in]{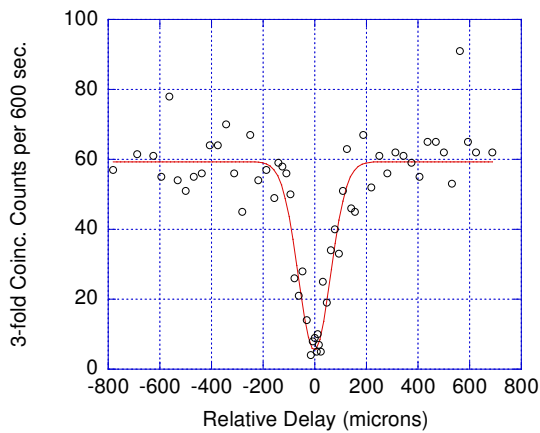}
\vspace*{-.25in}
\caption{Observation of a highly non-classical three-photon quantum interference dip \protect\cite{hong87,rarity97}.  The data shows the 3-fold coincidence counts (ie. gated two-photon events) as a function of the relative optical delay between  a heralded single-photon and a single photon post-selected from a weak coherent state.  The solid line is a least-squares fit to a simple Gaussian function with a visibility of ($90.8\pm 1.7$)\%. }
\label{fig:rtdip}
\end{figure}

\begin{figure}[t]
\vspace*{-.25in}
\hspace*{-.2in}
\includegraphics[angle=-90,width=3.5in]{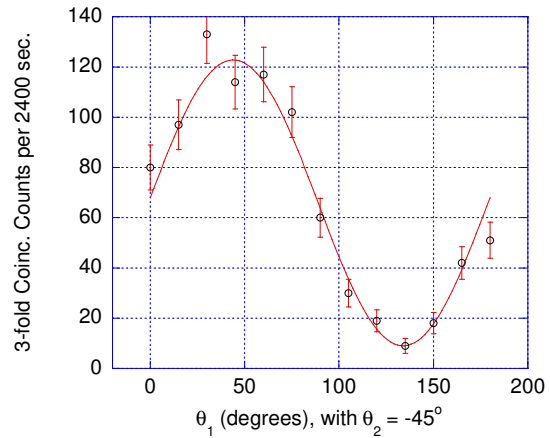}
\vspace*{-.25in}
\caption{Experimental signature of the polarization entangled state of equation (\protect\ref{eq:sa}) post-selected from independent sources. The data shows the accumulation of three-photon (ie. gated two-photon) coincidence counts as a function of the setting of $\theta_{1}$.  The solid line is a least squares fit to a Sine-squared function with a visibility of ($86.4 \pm 3.2$)\%.  This high visibility represents the main result of this paper.}
\label{fig:sacurve}
\end{figure}

The other requirement for high visibility in this experiment was the bandpass of the interference filters, which is primarily dictated by the dispersive properties of the down-conversion crystal, the pump pulse duration, and the crystal length \cite{digiuseppe97,grice97,keller97}. With our 0.7mm thick BBO down-conversion crystal, we found interference filters with a 3nm FWHM to be sufficient.   With these filters in place we would typically obtain approximately 23 down-conversion coincidence detections per second between the triggering detector $D_{t}$ and either $D_{1}$ or $D_{2}$, while the singles counting rate in $D_{t}$ was typically around 1300 counts per second.  This implied a value of $H \approx 28$, which reflects a combination of inefficient coupling of the down-converted photons into the fibers, as well as fiber losses, limited polarizer and filter transmission, and limited detection efficiency.  For the 76 MHz repetition rate of the mode-locked laser, this required us to limit the magnitude of the coherent state so that the singles counting rates (due to path $C$) in $D_{1}$ or $D_{2}$ were much less than $1.4 \times 10^{6}$ per second.  For the data shown below, we therefore kept this value at roughly $1.5\times 10^{5}$ counts per second ($\alpha \approx 4\times 10^{-3}$). Based on the signal-to-noise ratio obtained from these values, we would expect at most a 95\% visibility of the quantum interference patterns.

The data shown in Figure \ref{fig:rtdip} is a plot of the three-photon coincidence counting rate as a function of the optical delay imposed by the glass wedges.  In analogy with the Hong-Ou-Mandel experiment \cite{hong87}, the half-wave plate was used to horizontally polarize photons from the weak coherent state, and $\theta_{1}$ and $\theta_{2}$ were set to their horizontal values.  The experimental results show the expected gated Hong-Ou-Mandel dip \cite{hong87,rarity97} with a visibility of ($90.8 \pm 1.7$)\%.  As described above, this high visibility implied that the experimental conditions necessary to maintain the indistinguishability of the two photons were met when the relative optical delay corresponded to the bottom of the dip.

The test of Bell's inequality with photons from independent sources could then be implemented by vertically polarizing the photons from the weak coherent state.  The data shown in Figure \ref{fig:sacurve} is a plot of the number of gated two-photon coincidence detections in this situation as a function of the setting of the polarization analyzer $\theta_{1}$, with $\theta_{2}$ fixed at $-45^{o}$. Figure \ref{fig:sacurve} represents the main result of this paper. The experimental results show the functional $Sin^{2}(\theta_{1}-\theta_{2})$ signature of the entangled state in equation (\ref{eq:sa}), with a visibility of ($86.4 \pm 3.2$)\% \cite{poltest}.  As is well known, a visibility greater than 71\% in this situation is sufficient for a violation of variants of Bell's inequalities subject to certain reasonable assumptions \cite{clauser78}. 

We also gathered extensive data at the various combinations of $\theta_{1}$ and $\theta_{2}$ settings required for a test of the Clauser-Horne-Shimony-Holt (CHSH) version of the Bell inequality, for which local hidden-variables models are bound by a parameter $|S| \leq 2$ \cite{clauser69}. A discussion of the various assumptions and loopholes related to this experiment is beyond the intended scope of 
this paper \cite{settings}. We obtained an experimental value of $S = -2.44 \pm 0.13$, which is consistent with what one would expect from the visibility of 86\% observed in Figure \ref{fig:sacurve}.

Although a common laser beam was used throughout the experiment, the same effects would be expected theoretically if two different pump lasers had been used.  The two photons of interest were known to have been emitted in two different crystals, one in the Ti:Sapphire crystal of the laser and the other in the BBO down-conversion crystal.  Any potential phase relationship between these two sources should factor out of the final state in equation (\ref{eq:sa}) and have no effect on the results.  The absence of any coherent effects from the laser could be seen experimentally from the fact that the relevant beams propagated over long unstabilized paths that were known to cause large phase drifts on time scales much shorter than the data accumulation time.  In addition to the lack of any phase correlation, a comparison of the singles and coincidence rates showed that any intensity correlations between the two photon sources were less than 1\% and consistent with zero.   

In conclusion, we have violated Bell's inequality using one photon from parametric down-conversion and a second photon from an attenuated laser beam.  The two photons were entangled at a beam splitter through the post-selection technique of Shih and Alley \cite{shih88}. These results demonstrate that non-classical interference effects can be obtained using photons from independent sources, which is an important requirement for optical approaches to quantum information processing \cite{knill01,franson02}. 

We acknowledge useful discussions with M.M. Donegan, M.J. Fitch, and B.C. Jacobs.  This work was supported by ARO, NSA, ARDA, ONR, and IR\&D funding.




\begin{thebibliography}{50}

\bibitem{bellbook} {\em Speakable and unspeakable in quantum mechanics}, J.S. Bell, Cambride University Press, Cambridge (1987).

\bibitem{aspect81} A. Aspect, P. Grangier, and G. Roger, Phys. Rev. Lett. {\bf 47}, 460 (1981); {\bf 49}, 91 (1982).

\bibitem{kwiat95} P.G. Kwiat {\em et. al.} Phys. Rev. Lett. {\bf 75}, 4337 (1995).

\bibitem{yurke92} B. Yurke and D. Stoler, Phys. Rev. Lett. {\bf 68}, 1251 (1992); Phys. Rev. A {\bf 46}, 2229 (1992)

\bibitem{jennewein02} T. Jennewein, G. Weihs, J.-W. Pan, and A. Zeilinger, Phys. Rev. Lett. {\bf 88}, 017903 (2002).

\bibitem{zhao03} Z. Zhao, T. Yang, A.-N. Zhang, and J.-W. Pan, quant-ph/0301118.

\bibitem{zukowski93} M. Zukowski, A. Zeilinger, M.A. Horne, and A.K. Ekert, Phys. Rev. Lett. {\bf 71}, 4287 (1993).

\bibitem{shih88} Y.H. Shih and C.O. Alley, Phys. Rev. Lett. {\bf 61}, 2921 (1988).

\bibitem{knill01} E. Knill, R. Laflamme, and G.J. Milburn, Nature {\bf 409}, 46
(2001).

\bibitem{franson02} J.D. Franson, M.M. Donegan, M.J. Fitch, B.C. Jacobs, and T.B. Pittman, Phys. Rev. Lett. {\bf 89}, 137901.

\bibitem{feynmanlectures} {\em The Feynman Lectures on Physics, Vol. 3}, R.P. Feynman, R.B. Leighton, and M. Sands, Addison-Wesley (1965).

\bibitem{reid86} M.D. Reid and D.F. Walls, Phys. Rev. A {\bf 34}, 1260 (1986).

\bibitem{hong86} C.K. Hong and L. Mandel, Phys. Rev. Lett. {\bf 56}, 58 (1986).

\bibitem{ou97} Z.Y. Ou, Quantum Semiclass. Opt. {\bf 9}, 599 (1997).

\bibitem{koashi96} M.Koashi, M.Matsuoka, and T. Hirano, Phys. Rev. A {\bf 53}, 3621 (1996).

\bibitem{kuzmitch00} A. Kuzmitch, I.A. Walmsley, and L. Mandel, Phys. Rev. Lett. {\bf 85}, 1349 (2000).

\bibitem{resch02} K.J. Resch, J.S. Lundeen, and A.M. Steinberg, Phys. Rev. Lett. {\bf 87}, 123603 (2001); {\bf 88}, 113601 (2002); {\bf 89}, 037904 (2002).
 

\bibitem{lvovsky} A.I. Lvovsky, {\em et. al.}, Phys. Rev. Lett. {\bf 87}, 050402 (2001); {\bf 88}, 250401 (2002); Phys. Rev. A {\bf 65}, 033830 (2002); {\bf 66}, 011801R (2002). 


\bibitem{rarity97} J.G. Rarity and P.R. Tapster, Philos. Trans. R. Soc. London {\bf A 355}, 2267 (1997).

\bibitem{hong87} C.K. Hong, Z.Y. Ou, and L. Mandel, Phys. Rev. Lett. {\bf 59}, 2044 (1987).

\bibitem{zukowski95} M. Zukowski, A. Zeilinger, and H. Weinfurter, Ann. N.Y. Acad. Sci. {\bf 755} 91, (1995).

\bibitem{bouwmeester} D. Bouwmeester, J.-W. Pan, K. Mattle, H. Weinfurter, and A. Zeilinger, Nature {\bf 390}, 575 (1997); Phys. Rev. Lett. {\bf 82}, 1345 (1999).

\bibitem{pan98} J.-W. Pan, D. Bouwmeester, and A. Zeilinger, Phys. Rev. Lett. {\bf 80}, 3891 (1998).

\bibitem{ou99} Z.Y. Ou, J.-K. Rhee, and L.J. Wang, Phys. Rev. Lett. {\bf 83}, 959 (1999).

\bibitem{linares01} A. Lamas-Linares, J.C. Howell, and D. Bouwmeester, Nature {\bf 412}, 887 (2001).

\bibitem{pan01} J.-W. Pan, M. Daniell, S. Gasparoni, G. Weihs, and A. Zeilinger, Phys. Rev. Lett. {\bf 86}, 4435 (2001).

\bibitem{deried02} H. de Riedmatten, I. Marcikic, W. Tittle, H. Zbinden, and N. Gisin, quant-ph/0208174.

\bibitem{yam03} T. Yamamoto, M. Koashi, S.K. Ozdemir, and N. Imoto, Nature {\bf 421}, 343 (2003).

\bibitem{digiuseppe97} G. Di Giuseppe, L. Haiberger, F. De Martini, and A.V. Sergienko, Phys. Rev. A {\bf 56}, R21 (1997).

\bibitem{grice97} W.P. Grice and I.A. Walmsley, Phys. Rev. A {\bf 56}, 1627 (1997); Phys. Rev. A {\bf 64}, 063815 (2001).

\bibitem{keller97} T.E. Keller and M.H. Rubin, Phys. Rev. A {\bf 56}, 1534 (1997).


\bibitem{poltest} Repeating the experiment with $\theta_{2}$ fixed at $0^{o}$ showed the expected $Sin^{2}(\theta_{1})$ pattern indicitive of one vertically polarized photon and one horizontally polarized photon.

\bibitem{clauser78} J.F. Clauser and A. Shimony, Rep. Prog. Phys. {\bf 41}, 1881 (1978).

\bibitem{clauser69} J.F. Clauser, M.A. Horne, A. Shimony, and R.A. Holt, Phys. Rev. Lett. {\bf 23}, 880 (1969).

\bibitem{settings} Descriptions of the relevant polarizer settings and data accumulation techniques can be found, for example, in refs. \protect\cite{kwiat95} and \protect\cite{jennewein02}.


\end{thebibliography}
\end{document}